# Ballistic transport of long wavelength phonons and thermal conductivity accumulation in nanograined silicon-germanium alloys

Long Chen,[1] Jeffrey L. Braun,[2] Brian F. Donovan,[2,3] Patrick E. Hopkins,[1,2,4,*] and S. Joseph Poon[1,**]

[1] *Department of Physics, University of Virginia, Charlottesville, Virginia 22904-4714*

[2] *Department of Mechanical & Aerospace Engineering, University of Virginia, Charlottesville, Virginia 22904-4746*

[3] *Current Address: Department of Physics, United States Naval Academy, Annapolis, Maryland 21402-5002*

[4] *Department of Materials Science and Engineering, University of Virginia, Charlottesville, Virginia 22904-4745*

Computationally efficient modeling of the thermal conductivity of materials is crucial to thorough experimental planning and theoretical understanding of thermal properties. We present a modeling approach in this work that utilizes frequency-dependent effective medium to calculate lattice thermal conductivity of nanostructured solids. The method accurately predicts a significant reduction in the thermal conductivity of nanostructured $Si_{80}Ge_{20}$ systems, along with previous reported thermal conductivities in nanowires and nanoparticles-in-matrix materials. We use our model to gain insight into the role of long wavelength phonons on the thermal conductivity of nanograined silicon-germanium alloys. Through thermal conductivity accumulation calculations with our modified effective medium model, we show that phonons with wavelengths much greater than the average grain size will not be impacted by grain boundary scattering, counter to the traditionally assumed notion that grain boundaries in solids will act as diffusive interfaces that will limit long wavelength phonon transport. This is further supported through a modulation frequency dependent thermal conductivity as measured with time-domain thermoreflectance.

*phopkins@virginia.edu
**sjp9x@virginia.edu

The conflict between growing demands for energy and limited non-renewable energy sources has attracted large concern over the past few decades, which has spurred a multitude of researchers to explore clean and renewable energy. Thermoelectric (TE) materials, which can generate electricity from waste heat, could play an important role in a global sustainable energy solution. The performance of a thermoelectric material is evaluated by a dimensionless figure of merit ZT, which is equal to $S^2\sigma T/\kappa$ where $S$ is Seebeck coefficient, $\sigma$ is the electrical conductivity, and κ is the thermal conductivity. Among the TE materials, silicon-germanium structures continue to be the main focus of tremendous investment due to their widespread integration in TE power generation, optoelectronic devices, and high-mobility transistors. In these materials, it has been shown that the thermal conductivity, κ, can be decreased while preserving the electronic power factor, thus the figure of merit ZT is increased at much lower cost [1].

To simulate the lattice thermal conductivity of materials and nanosystems, several approaches have been advanced in the literature, including: the Callaway model [2,3] derived from the Boltzmann Transport Equation (BTE) [4-7], Monte Carlo simulations with varying phonon frequency dependence [8,9], and various methods to calculate the phonon mean free path (MFP) distributions, including analytical models, numerical results from molecular dynamics (MD) simulations [10], and first-principles calculations based on density functional theory [11]. Notably, Minnich and Chen used an Effective Medium Approach (EMA) to predict the thermal conductivity of heterogeneous nanostructures, and modified the phonon mean free paths in both matrix materials and nanoparticles. This modification came from assuming phonon single particle scattering off embedded nanoinclusions, as well as incorporating a thermal boundary resistance at the nanoinclusion/matrix interface. However, this method is only applicable in small volume fractions of embedded particles since it is based on first-order T-matrix approximation [12].

In this current work, we revisit the use of the MFP spectrum in nanocomposites and upgrade the effective medium approach [13,14] to include a phonon wavelength dependence (i.e., non-gray approach) in deriving the expressions for the lattice thermal conductivity of a bulk system; we then extend this derivation to their fully nanostructured states. We use our model to gain insight into the role of long wavelength phonons on the thermal conductivity of nanograined silicon-germanium alloys via measurements of the frequency dependence in thermal conductivity with time domain thermoreflectance (TDTR) [15-17]. Through thermal conductivity accumulation calculations with our modified effective medium model, we show that phonons with wavelengths much greater than the average grain size will not be impacted by grain boundary scattering, counter to the traditionally assumed notion that grain boundaries in solids will act as diffusive interfaces that will limit long wavelength phonon transport.

To develop our model, we start with the phonon Boltzmann transport equation:

$$\frac{\partial N_{q\lambda}}{\partial t} + \overrightarrow{v_{q\lambda}} \cdot \nabla N_{q\lambda} + \vec{F} \cdot \nabla_v N_{q\lambda} = \left(\frac{\partial N_{q\lambda}}{\partial t}\right)_c \quad (1)$$

where $N_{q\lambda}$ is the phonon distribution with wavenumber q, in phonon branch λ, $\overrightarrow{v_{q\lambda}}$ is the phonon group velocity, $\vec{F}$ is an external force, and $\left(\frac{\partial N_{q\lambda}}{\partial t}\right)_c$ is the time rate of change of the phonon distribution due to collisions. The fundamental assumptions of this work are: all of the parameters in Eq. (1) are phonon frequency dependent, the system is in an equilibrium state $\frac{\partial N_{q\lambda}}{\partial t} = 0$, and there is no external force. Thus, under the time relaxation approximation, we can derive heat flux in wave vector space as:

$$\vec{j} = -\sum_\lambda \iiint \hbar\omega_{q\lambda} (\vec{v_{q\lambda}} \cdot \nabla T \frac{\partial n_{q\lambda}}{\partial T} \tau_{q\lambda}) \vec{v_{q\lambda}} dq_i dq_j dq_k \quad (2)$$

Comparing this with $\vec{j} = -\kappa \nabla T$, we can write κ in q space as:

$$\kappa = \sum_\lambda \iiint \hbar\omega_{q\lambda} \frac{\partial n_{q\lambda}}{\partial T} \tau_{q\lambda} (v_{q_i\lambda})^2 \, dq_i dq_j dq_k \quad (3)$$

To determine the scalar form of the thermal conductivity, the spectral heat capacity and the tensor form of the thermal conductivity derived above are compared with $\kappa(q) = \frac{1}{3}C(q)v(q)L(q)$, where L(q) is the mean free path with wavenumber q. The spectral thermal conductivity for a given wavenumber q can be written as:

$$\kappa(q) = \frac{D}{(2\pi)^3} 4\pi q^2 \sum_\lambda \hbar\omega_{q\lambda} \frac{\partial n_{q\lambda}}{\partial T} (v_{q_\lambda})^2 \tau_{q\lambda} \quad (4)$$

where we use D as a normalization factor which is used to preserve the number of phonon modes of the system. We determine it by equating the experimental specific heat to the calculated value in parentheses in Eq. (4). The total phonon scattering time is determined by contributions from defect scattering, Umklapp scattering, and grain boundary scattering, and the forms of these scattering times and additional assumptions are outlined in the Supplemental Material [18]. Modeling the thermal conductivity of nanomaterials with the use of a fixed length scale, *l*, for boundary scattering, such as a film thickness or grain size, is common in literature, and assumes that all phonons with mean free paths greater than *l* will scatter at this boundary and have a limited contribution to heat transport compared to their bulk counterparts. We refer to this as the "Fixed Boundary Length" (FBL) model.

Moving beyond the FBL, it is important to account for the physical geometry of the structure that creates the boundary that could impede phonon transport. For example, the geometry, characteristic structure, and properties of the material that are creating the boundaries must also be accounted for in thermal conductivity calculations. Thus, we can consider boundary scattering, *l*, as the length characteristic of the material defined as $l = (\rho * \sigma)^{-1}$, where ρ is the number of particles in a unit volume, and σ is cross section. We can further break down σ as the combination of scattering cross section in two extreme regimes [19]:

$$\sigma_{total}^{-1} = \sigma_{Ray}^{-1} + \sigma_{nGeo}^{-1} \quad (5)$$

Where $\sigma_{Ray}$ and $\sigma_{nGeo}$ are scattering cross sections in the Rayleigh limit and near geometric limit, respectively. Kim and Majumdar proposed an approximate analytical solution to estimate the phonon scattering cross section of polydispersed spherical nanoparticles [19], which we adopt in our model calculations here and detail in the Supplemental Material [18]. We refer to our model calculations using $l = (\rho * \sigma)^{-1}$ with Eqs. (5) as the "Spectral Boundary Length" (SBL) model.

Finally, following Nan *et al*. [14,20] to extend our simulation to nano-structured systems, the differential effective medium approach (DEM) was applied. The advantage of DEM is that it uses pre-determined parameters when compared to other nearly parameter free methods. In addition, the modified term in DEM already accounts for the lower order inter-particle phonon scattering, which yields an easier and more efficient way to calculate lattice thermal conductivity. To apply the DEM, Poon *et al*. reformulated what Bruggeman developed from a physical standpoint and interpreted the results in terms of higher-order (multi-particle) scattering [21,22]. The key point is that at volume fraction $\varphi$, the thermal conductivity of the matrix is updated to $\kappa(\varphi)$. Previously, $d\varphi$ was substituted by an effective $d\varphi' = d\varphi/(1-\varphi)$, where $1 - d\varphi$ is the volume of the unoccupied host. By doing this, a small volume fraction of particles is incrementally added to the matrix. After each addition, the host is updated, and the added particle scatters phonons in the updated matrix, which gives rise to the multi-particle effect. Upon addition of the particles in the matrix up to a particle volume fraction, $\varphi$, approaching 100%, we achieve the lattice thermal conductivity of a specific phonon with wavenumber q. This can be expressed as:

$$\kappa(q) = \frac{\kappa_p(q)}{1+\frac{\alpha_{p0}(q)*\kappa_p(q)}{\kappa_{p0}(q)}} \quad (6)$$

where, $\kappa_p(q)$ is the thermal conductivity of embedded nanoparticle with wavenumber q, $\kappa_{p0}(q)$ is the intrinsic lattice thermal conductivity of particle material in bulk form, and thermal resistance parameter $\alpha_{p0} = R_{p0}\kappa_{p0}/(d/2)$, where $R_{p0}$ is the thermal barrier resistance defined as $R_{p0} \approx 8/(C_{p0}v_{p0})$[23]. We refer to this model as the "DEM" model; more details of this model are provided in the Supplemental Material [18].

To validate our model, we compare our model calculations to measured thermal conductivity of silicon nanowires by Li *et al* [24]. As shown in Fig. 1(a), our model can also capture the thermal conductivity reduction in Si nanowires due to boundary scattering, similar to more traditional phonon-transport models based on semi-classical formalisms, as summarized by Yang and Dames [24]. We also calculate the thermal conductivity of a Si/Ge nanocomposites (*d* = 10 nm Si nanoparticles in a Ge bulk matrix) and compare these calculations against various previous models in Fig. 1(b). Our non-gray DEM simulation results in more rapid reductions in the thermal conductivity as a function of volume fraction at low volume fractions as compared with Minnich's gray and non-gray EMA models [12]. Also our non-gray DEM results agree well with Jeng's Monte Carlo (MC) simulations, which are supported by experimental results [25-27]. In addition, Miranda *et al.* [28] showed that DEM and finite element method were consistent to each other in their thermal conductivity study.

Taking the results in Fig. 1(a) and (b) together, the use of our modified DEM in our method for phonon thermal conductivity predictions yields an easier, more efficient, and more accurate way to calculate the lattice thermal conductivity of both homo- and heterogeneous nanomaterials and nanocomposites as compared to previously used approaches. Therefore, our proposed model can be used to study phonon transport in nanocomposites systems going beyond simple thermal conductivity predictions via calculations of phonon thermal conductivity accumulation as a function of mean free path.

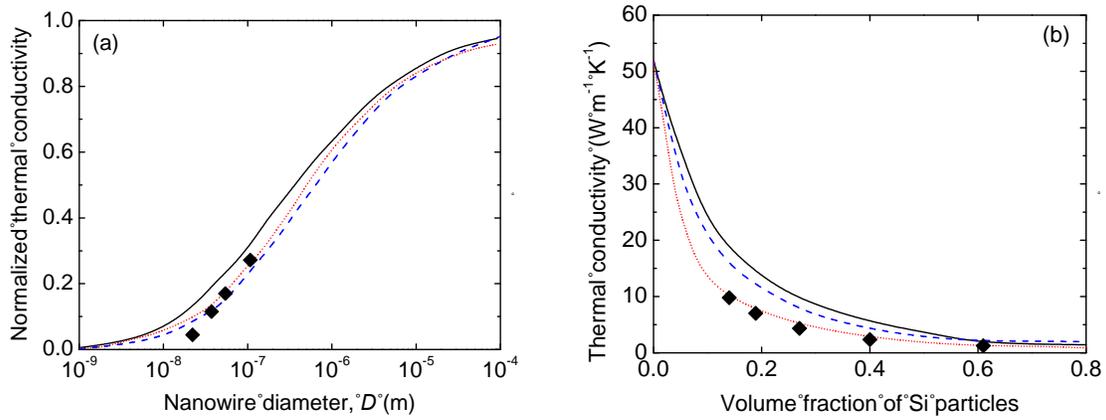

FIG. 1. (a) Normalized thermal conductivity of silicon nanowires calculated using the Holland model (solid, black) and BvKS model (dashed, blue), both of which were taken from Yang and Dames (Ref. [24]), as compared to our non-gray DEM (dotted, red), and the experimental data from Li *et al* (black diamonds). (b) Lattice thermal conductivity $\kappa_L$ of Si/Ge nanocomposite dependence on Si nanoparticles's volume fraction at room temperature (T=300K), with an average grain size $d = 10$ nm. Non-gray DEM simulation with grain size dispersion-standard deviation of $0.577d$ (dotted, red) is compared with Jeng's MC simulation (black diamonds), gray EMA (black, solid), and non-gray EMA (blue dashed).

In the remainder of this work, we extend the use of our DEM approach to nanograined Si-Ge bulk systems. We fabricate a silicon control sample and $Si_{80}Ge_{20}$ samples with varying grain sizes. Ingots of both compositions are prepared by arc melting under an Argon atmosphere. Ingots are pulverized into 1 to 30 μm size powders. Those micro powders are consolidated using Spark Plasma Sintering (Thermal Technologies SPS 10-4). To produce fully dense compacted disks with relative large grain sizes, Si and $Si_{80}Ge_{20}$ samples are sintered at 1280 °C and 1210 °C, repsetively, for 4 minutes under 60 MPa. As for the nanostructured systems $Si_{80}Ge_{20}$, micro-powders of Si and $Si_{80}Ge_{20}$ are loaded into a 440C stainless steel vial as well as two 0.5" and four 0.25" stainless steel balls. This process is performed in a glove box under Argon atmosphere. The vial is then sealed and placed in a SPEX 8000D vibrational mixer. The powders are ball milled for 40 hr for the $Si_{80}Ge_{20}$ systems. The ball milled powders are then compacted by SPS. We determine the grain size of the disks by cleaved cross section analyzed under SEM (Fig. 2). The grain size distribution is determined by SPIPTM$^{TM}$. The grain sizes for the silicon control extend to as large as 30 μm, while the ranges of grain sizes determined for the various $Si_{80}Ge_{20}$ samples are listed in the caption of Fig. 2. We mechanically polish all samples after deposition to facilitate TDTR measurements, and the resulting RMS roughnesses determined via mechanical profilometry maps were 30 ± 10 nm.

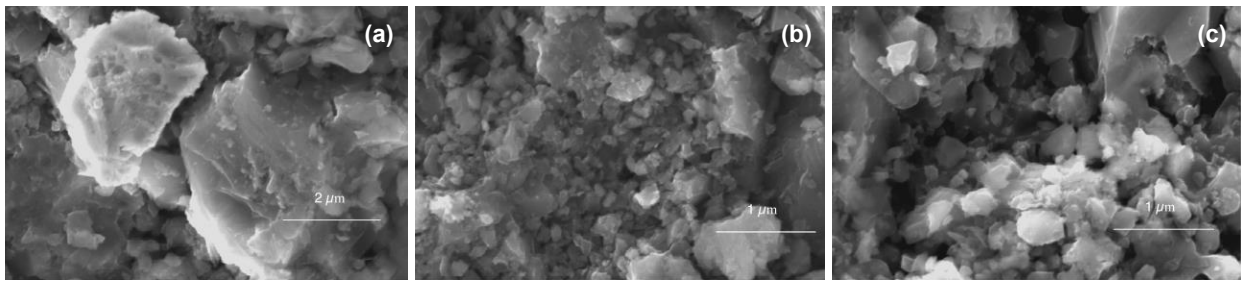

FIG. 2. SEM micrographs $Si_{80}Ge_{20}$ systems: (a) $Si_{80}Ge_{20}$ 2.0 ± 0.17 µm, (d) $Si_{80}Ge_{20}$ 110 ± 21 nm, and (e) $Si_{80}Ge_{20}$ 73 ± 29 nm.

We conduct experimental measurements of the thermal conductivity of these systems using TDTR with varying pump modulation frequencies, which effectively varies the heater length scale. This approach has recently been vetted for studying accumulation effects on the thermal conductivity for alloys [17]. The fabrication and characterization of the various Si and $Si_{80}Ge_{20}$ samples, and details of our TDTR measurements, including validity of the use of a diffusive heat equation-based model for TDTR data analysis [17], are included in the Supplemental Material [18].

We alter the modulation frequency of the pump beam during our TDTR measurements from 1.49 MHz to 12.2 MHz. We calculate the depth probed during the TDTR measurements using the heat diffusion equation based on twice the 1/e decay of the temperature from the surface of the substrate, consistent with the recent findings of Koh *et al*. [29]. The use of our relatively large pump and probe spots sizes allow our TDTR measurements at the various frequencies to be directly related to the thermal transport physics in the cross-plane direction, therefore reducing measurement sensitivities to in-plane non-diffusive thermal transport [17]. Furthermore, the use of Al as our thin film transducer will allow for direct comparison of our measurements to previous reports of Fourier failure in Si-Ge-based systems without the potential for additional electron-phonon resistances in the metal film to complicate our results and analyses [30].

As discussed by several recent works [17,30-33], we relate the changing thermal penetration depth with the varying frequency during a TDTR experiments to changes in the net heat flux. Thus, we modify our EMA modeling approach to separate the thermal conductivity of the samples into a high frequency mode component (diffusive) and low frequency mode component (quasi-ballistic). To determine the cut-off MFP that separates this high and low frequency mode regime, a plausible way is to equate the MFP to the thermal penetration depth, and then use this MFP to determine the corresponding wavenumber $q$. As another approach, we also use a compact heat conduction model based on the two-fluid assumption (bridge function) [33]. In this model, we write the net heat flux in form of: $J = j^{LF} + j^{HF}$, where $j^{LF}$ and $j^{HF}$ are the low and high frequency mode contributions to the heat flux. This approach is detailed in the Supplemental Material [18]. We find that for these alloys, both approaches yield similar results due to the alloy transport being dominated by the near-zone center modes.

We now use our approach to analyze the TDTR data taken on the $Si_{80}Ge_{20}$ systems, comprised of samples with average grain sizes of 2 µm, 110 nm, and 73 nm. For all nanograined samples, we observe a frequency dependence in the measured thermal conductivity, as shown in Fig. 3, implying that as our measurement depth is increased, the thermal conductivity is also increased. Note, these data show similar trends to previously reported frequency dependent TDTR data on SiGe alloys [17,29]. We attribute this effect to the accumulation of the lattice thermal conductivity as we probe into the length scales that capture the heat carrying mean free paths in this system. Based on previous works, this frequency dependent trend might suggest that phonons with mean free paths greater than the thermal penetration depth are carrying substantial amount of heat in these systems, and our measurements are related to an accumulation of phonon mean free paths [17,29,30]. This, however, is counter-intuitive to traditionally-implemented phonon transport dynamics that assume phonons with mean free paths greater than the grain size will scatter and thus not contribute to thermal conductivity (as predicted via the traditionally assumed FBL model).

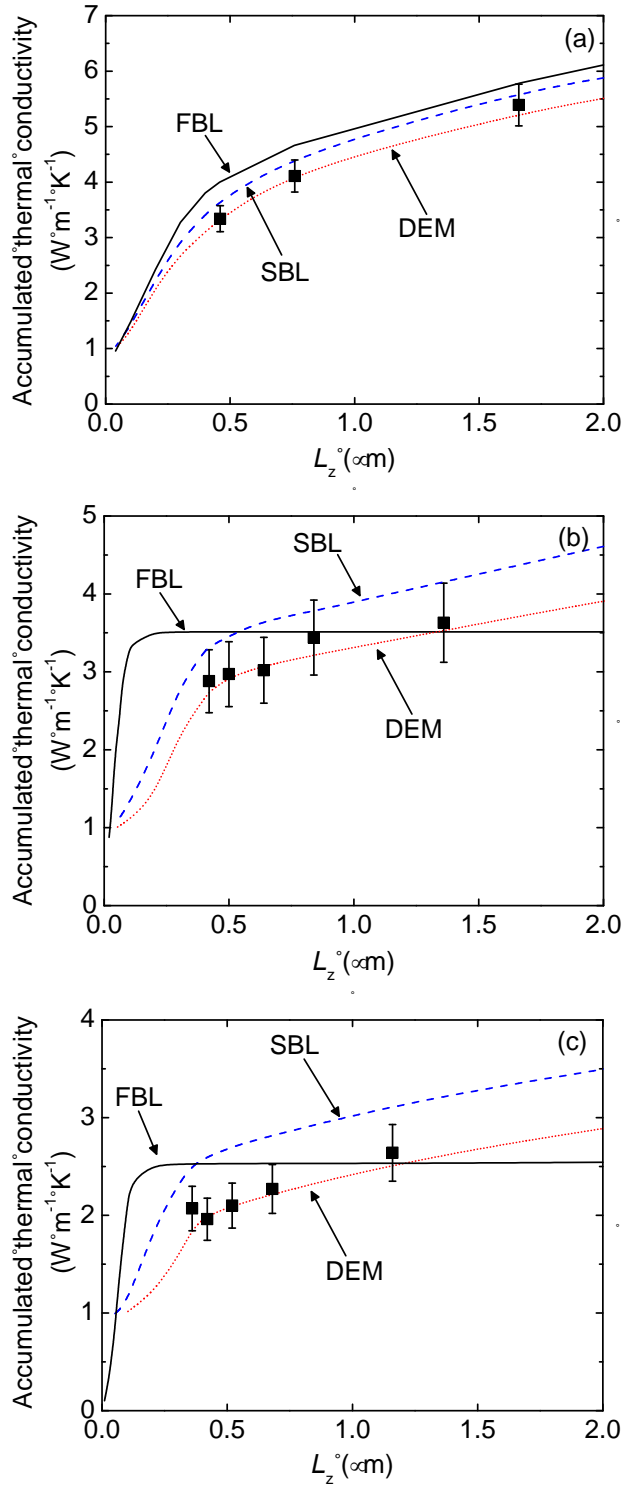

**FIG. 3. Modeled (lines) and measured (points) accumulated thermal conductivity vs. thermal penetration depth for $Si_{80}Ge_{20}$ samples with (a) average grain size 2 μm, (b) 110 nm, and (3) 73 nm.**

The traditionally implemented assumptions in the FBL model assume all phonons scatter at a length scale defined by the grain boundary. Thus, regardless of the assumption of what phonons are considered quasi-ballistic and diffusive (i.e., hard cutoff or bridge function to separate these two regimes), the FBL always predicts a leveling off of the accumulative thermal conductivity at much shorter length scales than that observed in the experimental data (Fig. 3), since by definition, it is restricting propagation of long wavelength phonons. The use of bridge function and implementation of the two-fluid assumption in the modeling of the accumulation of thermal conductivity yields predicted trends in good agreement with our measured frequency dependent thermal conductivity data. For the three different grain sizes, our DEM approach exhibits improved agreement with the experimental data over the varying thermal penetration depths as compared to the other models (FBL and SBL).

The FBL fails to account for the long wavelength phonon transport, as it assumes these phonons will scatter with the grain boundaries. This discrepancy is most pronounced for the 110 and 73 nm samples. While long wavelength-dominated phonon transport is well known in crystalline alloys (due to high frequency phonon-mass impurity scattering) [34,35], the role of grain boundaries and their interplay among long wavelength phonon transport and phonon-mass impurity scattering has been less-frequently explored. Our results suggest that creating nanograins in crystalline alloys may not have as pronounced as an effect on lowering thermal conductivity as predicted by traditional boundary scattering theories, such as those assumed in the FBL. Since the majority of the thermal transport in crystalline alloys is driven by the long wavelength, long wave vector phonons (c.f., Fig. 3), the scattering cross section of nanograins could be too small to create significant impact on the majority of the heat carrying phonons, which is demonstrated by the modest frequency dependence in the 110 and 73 nm alloys compared to the 2 μm sample.

In summary, we present a modeling approach that utilizes a frequency-dependent effective medium method to calculate the lattice thermal conductivity of nanostructured solids. It is the implementation of frequency dependence to DEM for the first time. It is the first time that the frequency-dependent model is implemented to DEM. And with the application of matching high and low phonon regimes to determine the cut-off MFP, we gain insight into the role of long wavelength phonons on the thermal conductivity of nanograined silicon-germanium alloys. Through thermal conductivity accumulation calculations with our modified effective medium model, we show that phonons with wavelengths much greater than the average grain size will not be impacted by grain boundary scattering, counter to the traditionally assumed notion that grain boundaries in solids will act as diffusive interfaces that will limit long wavelength phonon transport.

This material is based upon work supported by the Air Force Office of Scientific Research under Award No. FA9550- 15-1-0079.

# SUPPORTING INFORMATION

## S1. Details of the "Fixed Boundary Length" (FBL) and "Spectral Boundary Length" (SBL) models, assumptions and calculations

In this work, when modeling silicon, we assume a polynomial fit to literature data [1]. For the SiGe calculations in this work, we assume a sine-type phonon distribution, so that the group velocity is given by $v_q = v_s \cos(qa/2)$ and the frequency $\omega_q = \omega_{max} \sin(qa/2)$, where $v_s$ is the sound speed at $q = 0$ and $a$ is the lattice constant. $v_s$, $\omega_{max}$, and $a$ are 5166 m/s, 6.42 THz, and 5.472 Å.

The total phonon scattering time is determined by contributions from defect scattering, Umklapp scattering, and grain boundary scattering. These scattering mechanisms are combined using Matthiessen's rule: $\tau_{total}(q)^{-1} = \tau_{defect}(q)^{-1} + \tau_{umkl}(q,T)^{-1} + \tau_b(q)^{-1}$ Each scattering component is modeled as follow:

$$\tau_{defect}(q)^{-1} = A\omega^4 \qquad (S1)$$

$$\tau_{umkl}(q,T)^{-1} = B\omega^2 T e^{(-C/T)} \qquad (S2)$$

$$\tau_b(q)^{-1} = v/l \qquad (S3)$$

The constants A, B and C are determined by fitting these to the experimental data. $A = 2.41 * 10^{-45}\ s^3$, $B = 4.13 * 10^{-19}\ s\ K^{-1}$, and $C = 4.13 * 151\ K$ are found to be the best fit parameters.

It is important to account for the physical geometry of the structure that creates the boundary that could impede phonon transport. For example, circular boundaries could impact phonon scattering differently than rectangular boundaries. Furthermore, the characteristic structure and properties of the material that is creating the boundaries and structures in a host material must also be accounted for in thermal conductivity calculations (e.g., mass and force constant changes at, for example, a grain or nanoparticle boundary), we can consider boundary scattering, $l$ as the length characteristic of the material defined as $l = (\rho * \sigma)^{-1}$, where $\rho$ is the number of particles in a unit volume which can be determined by $\rho = 6\varphi/\pi R^3$, where R is the radius of the spherical nanoparticle, and σ is cross section. We can further break down σ as the combination of scattering cross section in two extreme regimes [2]:

$$\sigma_{total}^{-1} = \sigma_{Ray}^{-1} + \sigma_{nGeo}^{-1} \qquad (S4)$$

Where $\sigma_{Ray}$ and $\sigma_{nGeo}$ are scattering cross sections in the Rayleigh limit and near geometric limit, respectively.

Kim and Majumdar proposed an approximate analytical solution to estimate the phonon scattering cross section of polydispersed spherical nanoparticles. They perturbed the Hamiltonian via differences in mass and bond stiffness between a host medium and a spherical nanoparticle for the scattering cross section in the Rayleigh limit, using the van de Hulst approximation for anomalous diffraction in the near geometrical limit [2].

Based on Kim and Majumdar's work, the scattering cross sections at the Rayleigh limit and near geometrical limit can be expressed as:

$$\sigma_{Ray} = \pi R^2 \chi^4 \left( \frac{\beta^2}{4}\left(\frac{\Delta M}{M}\right)^2 + 3\beta^8 \left(\frac{\Delta K}{K}\right)^2 \frac{(\sin\frac{\beta|\vec{q}|\delta}{2})^4}{\left(\frac{\beta|\vec{q}|\delta}{2}\right)^4} \right) \frac{\pi(\cos(4\chi)-1+(4\chi)\sin(4\chi)+32\chi^4-8\chi^2)}{16\chi^6} \quad (S5)$$

$$\sigma_{Geo} = 2\pi R^2 \left(1 - \frac{\sin\left(2\chi\left(\frac{q'}{q}-1\right)\right)}{\chi\left(\frac{q'}{q}-1\right)} + \frac{\left(\sin\left(\chi\left(\frac{q'}{q}-1\right)\right)\right)^2}{\left(\chi\left(\frac{q'}{q}-1\right)\right)^2}\right) \quad (S6)$$

Where $\chi$ is the size parameter, defined as $\chi = qR$, where q is the incoming wave vector, q' is the scattered wave vector, $\delta$ is the volume size, and $\beta$ is the polar angle; M and $\Delta M$ are the mass of the host and the mass difference between the host and the spherical nanoparticle; K and $\Delta K$ denote the force constant of the host medium and the force constant difference between the host and the spherical nanoparticle. The values which we used in the calculation are $\Delta M/M = 0.326$, and $\Delta K/K = 0.131$ by considering the grain distribution.

To take the grain size dispersion into account, the effective scattering cross section was calculated with a normalized grain distribution function F(x) [3]:

$$\sigma_{eff} = \int_{d_{min}}^{d_{max}} \sigma_{total}(x) F(x) dx \quad (S7)$$

where $\sigma_{eff}$ is the effective scattering cross section, $\sigma_{total}(x)$ is the total scattering section, $d_{max}$ and $d_{min}$ are the upper limit and the lower limit of grain size, and F(x) is the normalized Gamma distribution defined as $F(x) = x^{a-1}\exp[-x/b]/b^a\Gamma(a)$, with the shape parameter a, and scale parameter b, which can be determined by fitting the grain size distribution. The parameters (a, b) are found to be (23.5, 0.085), (5.5, 0.2) and (18.25, 0.4) for $Si_{80}Ge_{20}$ 2.0 ± 0.17 μm, $Si_{80}Ge_{20}$ 110 ± 21 nm, and $Si_{80}Ge_{20}$ 73 ± 29 nm, respectively.

## S2. TDTR Measurements

We measure the thermal conductivity of a silicon control sample and $Si_{80}Ge_{20}$ with time domain thermoreflectance [4-6]. Time domain thermoreflectance (TDTR) utilizes a train of ultra-fast laser pulses to thermally stimulate a material system, and a time delayed probe pulse to measure the change in thermoreflectance due to the decay of the thermal energy deposited by the pump pulse. This work utilizes sub-picosecond laser pulses emanating from a Ti:Sapph laser system at 80 MHz. The probe pulses are monitored via lock-in detection at the pump modulation frequency for up to of 5.5 ns using a mechanical delay stage. We deposit a nominally 80 nm thin film of Al on the sample surface to act as a thermal transducer. We fit a thermal model to the decay to determine the thermal conductivity of the Si or $Si_{80}Ge_{20}$ samples. We assume literature values for the other physical properties included in the thermal model including the heat capacity of Si, $Si_{80}Ge_{20}$, and Al [7,8]. We assume a reduced thermal conductivity of the Al film based on electrical resistivity measurements and the Wiedemann-Franz law, though we are insensitive to this parameter in our experiment due to our spot sizes and pump-probe delay. We also treat the thermal boundary conductance for the Al/Si or Al/Si$_{80}$Ge$_{20}$ interface as a free parameter in our model fit. The analysis methods are described in greater detail elsewhere [4-6]. Reported error in the thermal conductivity measurement arises from small thickness variation of the Al transducer and measurement of a number of different sites on the surface of each sample. Using laser spot sizes of 50 μm and 17 μm diameters for the pump and probe, respectively, allows us to assume nearly one dimensional heat transfer in the through-plane direction [6]. The bulk Si control sample thermal conductivity was measured to be 135 ± 20 W m$^{-1}$ K$^{-1}$ and independent of modulation frequency.

### S3. The cut-off mean free path vs thermal penetration depth determined by the bridge function and hard cutoff approaches

We can state equations for the low and high frequency mode heat fluxes at wavenumber q as:

$$j(q)^{LF} = \frac{3}{5}(\text{MFP}(q))^2 \frac{\partial^2 j(q)^{LF}}{\partial x^2} - k(q)^{LF} \frac{\partial T}{\partial x} \quad (S8)$$

$$j(q)^{HF} = -k(q)^{HF} \frac{\partial T}{\partial x} \quad (S9)$$

Appling Eq. (4) in the main text and $\text{MFP}(q) = \tau_q v_q$ to the equations above, we can achieve:

$$j(q)^{LF} = \frac{3}{5}(\tau_q v_q)^2 \frac{\partial^2 j(q)^{LF}}{\partial x^2} - \left(\frac{D}{(2\pi)^3} 4\pi q^2 \hbar \omega_{q\lambda} \frac{\partial n_{q\lambda}}{\partial T}\right) v_q^2 \tau_q \frac{\partial T}{\partial x} \quad (S10)$$

$$j(q)^{HF} = -\left(\frac{D}{(2\pi)^3} 4\pi q^2 \hbar \omega_{q\lambda} \frac{\partial n_{q\lambda}}{\partial T}\right) v_q^2 \tau_q \frac{\partial T}{\partial x} \quad (S11)$$

The total heat-flux for both low frequency mode and high frequency mode contributions can be achieved by integrating $j(q)^{LF}$ and $j(q)^{HF}$ over corresponding $q$ space. We can find the wavenumber $q$ corresponding to cut-off MFP by varying $q$ until the total heat-flux in the low frequency mode equals to the total heat-flux in the high frequency mode; in other words, we make the assumption in this model that $j^{LF} = j^{HF}$, which defines the cutoff wavenumber. The cut-off MFP is determined by using this cutoff wavenumber $q$. The normalization factor D was found to be 0.901 by equating the experimental specific heat to the calculated value in the model.

The cut-off mean free path determined by hard cutoff equates the thermal penetration depth achieved by TDTR to the mean free path. Any phonons with mean free path larger than the measured thermal penetration depth are eliminated, and do not make any contribution to the accumulated thermal conductivity. The cut-off mean free path determined by bridge function separates the thermal conductivity of the samples into a high frequency mode component (diffusive) and low frequency mode component (quasi-ballistic). This method model equates the low and high frequency mode heat fluxes, which defines the cutoff wavenumber.

### S4. Validity of the use of a diffusive heat equation-based model when analyzing TDTR data

Wilson and Cahill [10] have questioned the validity of assuming a diffusive heat equation in analyzing TDTR data when frequency dependence in the thermal conductivity is observed. More specifically, as diffusion assumes a local temperature can be defined, if too many phonons are "escaping" the TDTR measurement volume defined from the spot size and $L_z$, then the majority of the phonons that define the heat capacity of the solid would not contribute to the measured temperature change, and thus these quantities (heat capacity and temperature) are ill-defined based on literature parameters. They note that to use traditionally implemented TDTR heat diffusion models [5,6,9] to analyze the experimental data, the majority of the phononic system that contributes to heat capacity must experience a diffusive scattering event in the measurement volume (i.e., phonons with $q > 0.1 q_{max}$, where for Si, for example, $q_{max} = 2\pi/a$). This type of transport would take place in a crystalline alloy, and lead to the frequency dependence in TDTR-measured thermal conductivity, since the majority of the heat conduction is driven by the long wavelength phonons due to the scattering of short wave vector, high frequency phonons with mass impurities [10-12]. In other words, in the TDTR measurements presented in this work, the Fourier law does not fail in the analysis of any of these samples [9]. The spectral contribution to thermal conductivity in the nanograined $Si_{80}Ge_{20}$ samples cannot be predicted from traditionally assumed boundary scattering models (e.g., FBL, which will truncate phonon transport at limiting length scale, such as a grain boundary). Phonon scattering cross sections, such as those calculated via Eqs. (S4)-(S7), must be accounted for to properly model this phenomenon.

The 1/e thermal penetration depth ($L_z$) was determined using the solution to the radially symmetric heat diffusion equation where the full spatial temperature profile was calculated [13]. The $L_z$ of interest is that within the $Si_{80}Ge_{20}$ substrate. As such, the temperature at the substrate surface adjacent to the Al transducer was used as the reference temperature for which the 1/e decay length was determined. Because the pump radius, $r_0$, was very large such that for all modulation frequencies ($\omega_0$) used, $\omega_0 \gg \mathfrak{D}/r_0^2$, where $\mathfrak{D}$ is the thermal diffusivity of $Si_{80}Ge_{20}$, the $L_z$ calculated was the same as that estimated using the equation

$$L_z = \sqrt{\frac{2\mathfrak{D}}{\omega_0}}$$

Koh *et al*. [14] showed that the amplitude of heat conduction at high frequencies can be approximated using an effective thermal conductivity reduced by omitting ballistic phonons with mean free paths greater than $2L_z$. In agreement with this finding, we use $2L_z$ to compare experimentally determined effective thermal conductivities of nanograined $Si_{80}Ge_{20}$ with the FBL, SBL, and DEM models presented in this work.

## S5. Details of spectral thermal conductivity calculations using the DEM approach for nanograined $Si_{80}Ge_{20}$ systems

To achieve the spectral thermal conductivity for nanograined systems, we add $d\varphi$ of particles to the host. After each addition, the thermal conductivity of the new host $\kappa_h(\varphi)$ is updated to $\kappa\left(\frac{\varphi}{1-d\varphi}\right)$. This updated term can be expanded to:

$$\kappa\left(\frac{\varphi}{1-d\varphi}\right) = \kappa(\varphi) + \varphi d\kappa(\varphi) \quad (S12)$$

So the transformation relation can be expressed as: $\kappa_h(\varphi) \to \kappa(\varphi) + \varphi d\kappa(\varphi)$

Then we consider the phonon scattering by the added particles, the updated thermal conductivity of the host becomes:

$$\kappa_h(q, \varphi) = \frac{1}{3}C_h(q, \varphi)v_h(q, \varphi)L_{h,x}(q, \varphi + d\varphi) = \frac{1}{3}C_h(q, \varphi)v_h(q, \varphi)^2(A(\omega_q)^4 + B(\omega_q)^2 T^3 + v\frac{6(\varphi + d\varphi)}{\pi a^3}(\sigma_{Rayleigh}^{-1} + \sigma_{near\ geometrical}^{-1})^{-1})^{-1} \quad (S13)$$

Expand the equation above and only keep the first-order terms, we can get:

$$\kappa_h(q, \varphi) = \kappa(q, \varphi)\left(1 - \frac{18\kappa(q,\varphi)(\sigma_{Rayleigh}^{-1} + \sigma_{near\ geometrical}^{-1})^{-1}}{\pi v(q,\varphi)a^3 C(q,\varphi)} d\varphi\right) \quad (S14)$$

By writing in terms of $d\varphi$, we can achieve that

$$\kappa(q, \varphi) = \kappa_{host}(q, \varphi)\left(1 + \frac{3(\kappa_p(q)(1-\alpha(q,\varphi)) - \kappa_{host}(q,\varphi))}{\kappa_p(q)(1+2\alpha(q,\varphi)) + 2\kappa_{host}(q,\varphi)} d\varphi\right) \quad (S15)$$

In addition, we have

$$\kappa(q, \varphi + d\varphi) = \kappa(q, \varphi) + d\kappa(q, \varphi) \quad (S16)$$

Combining (4) and (5), we can achieve that when $\varphi$ is approaching 1:

$$\kappa(q) = \frac{\kappa_p(q)}{1 + \frac{\alpha_{p0}(q) * \kappa_p(q)}{\kappa_{p0}(q)}} \quad (S17)$$

$\kappa_p(q)$ is the thermal conductivity of embedded nanoparticle with wavenumber q, defined as:

$$\kappa_p(q) = \frac{1}{3}C_p(q)v_p(q)L_p(q) \quad (S18)$$

where,

$$C_p(q) = 9\frac{9N}{(q_m)^3}q^2\frac{\hbar(qv)^2}{k_\beta T^2}\frac{e^{\frac{\hbar qv}{k_\beta T}}}{(e^{\frac{\hbar qv}{k_\beta T}}-1)^2} \quad (S19)$$

$$L_p(q) = v_p(q) * \left(A(\omega_q)^4 + B\omega_q^2 Te^{(-c/T)} + v_s/l\right)^{-1} \quad (S20)$$

$\kappa_{p0}(q)$ is the intrinsic lattice thermal conductivity of particle material in bulk form, defined as:

$$\kappa_{p0}(q) = \frac{1}{3}C_p(q)v_p(q)L_{p0}(q) \quad (S21)$$

$$L_{p0}(q) = v_p(q) * \left(A(\omega_q)^4 + B\omega_q^2 T e^{(-c/T)}\right)^{-1} \quad (S22)$$

and the thermal resistance parameter $\alpha_{p0} = R_{p0}\kappa_{p0}/(d/2)$, where $R_{p0} \approx 8/(C_{p0}v_{p0})$.

Further specific details of each of the three model calculations are listed below

*Fixed Boundary Length (FBL) Model*:
$$\kappa_p(q) = \frac{1}{3}C_p(q)v_p(q)L_p(q)$$

$$C_p(q) = 9\frac{9N}{(q_m)^3}q^2\frac{\hbar(qv)^2}{k_\beta T^2}\frac{e^{\frac{\hbar qv}{k_\beta T}}}{(e^{\frac{\hbar qv}{k_\beta T}}-1)^2}$$

$$L_p(q) = v_p(q) * \left(A(\omega_q)^4 + B\omega_q^2 T e^{(-c/T)} + v_s/L\right)^{-1}$$

L is the particle size.

*Spectral Boundary Length Model*:
$$\kappa_p(q) = \frac{1}{3}C_p(q)v_p(q)L_p(q)$$

$$C_p(q) = 9\frac{9N}{(q_m)^3}q^2\frac{\hbar(qv)^2}{k_\beta T^2}\frac{e^{\frac{\hbar qv}{k_\beta T}}}{(e^{\frac{\hbar qv}{k_\beta T}}-1)^2}$$

$$L_p(q) = v_p(q) * \left(A(\omega_q)^4 + B\omega_q^2 T e^{(-c/T)} + v_s/l\right)^{-1}$$

where $l$ is the length characteristic, defined as $l = (\rho * \sigma)^{-1}$, and is frequency dependent.

$$\sigma_{Ray} = \pi R^2 \chi^4 \left(\frac{\beta^2}{4}\left(\frac{\Delta M}{M}\right)^2 + 3\beta^8\left(\frac{\Delta K}{K}\right)^2\frac{(\sin\frac{\beta|\bar{q}|\delta}{2})^4}{\left(\frac{\beta|\bar{q}|\delta}{2}\right)^4}\right)\frac{\pi(\cos(4\chi)-1+(4\chi)\sin(4\chi)+32\chi^4-8\chi^2)}{16\chi^6}$$

$$\sigma_{nGeo} = 2\pi R^2\left(1 - \frac{\sin\left(2\chi\left(\frac{q'}{q}-1\right)\right)}{\chi\left(\frac{q'}{q}-1\right)} + \frac{\left(\sin\left(\chi\left(\frac{q'}{q}-1\right)\right)\right)^2}{\left(\chi\left(\frac{q'}{q}-1\right)\right)^2}\right)$$

*Differential Effective Medium (DEM) Model*:

$$\kappa(q) = \frac{\kappa_p(q)}{1+\frac{\alpha_{p0}(q)*\kappa_p(q)}{\kappa_{p0}(q)}}$$

where $\kappa_p(q) = \frac{1}{3}C_p(q)v_p(q)L_p(q)$

$$C_p(q) = 9\frac{9N}{(q_m)^3}q^2\frac{\hbar(qv)^2}{k_\beta T^2}\frac{e^{\frac{\hbar qv}{k_\beta T}}}{(e^{\frac{\hbar qv}{k_\beta T}}-1)^2}$$

$$L_p(q) = v_p(q) * \left(A(\omega_q)^4 + B\omega_q^2 Te^{(-c/T)} + v_s/l\right)^{-1}$$

$\kappa_{p0}(q)$ is the intrinsic lattice thermal conductivity of particle material in bulk form, defined as:

$$\kappa_{p0}(q) = \frac{1}{3}C_p(q)v_p(q)L_{p0}(q)$$

$$L_{p0}(q) = v_p(q) * \left(A(\omega_q)^4 + B\omega_q^2 Te^{(-c/T)}\right)^{-1}$$

and the thermal resistance parameter $\alpha_{p0} = R_{p0}\kappa_{p0}/(d/2)$, where $R_{p0} \approx 8/(C_{p0}v_{p0})$.